# On the Performance of Transmit Antenna Selection Based on Shadowing Side Information⋆

Ahmet Yılmaz, Ferkan Yılmaz, Mohamed-Slim Alouini and Oğuz Kucur

*Abstract*—In this paper, a transmit antenna selection scheme, which is based on shadowing side information, is investigated. In this scheme, the selected single transmit antenna provides the highest shadowing coefficient between transmitter and receiver. By the proposed technique, the frequency of the usage of the feedback channel from the receiver to the transmitter and also channel estimation complexity at the receiver can be reduced. We study the performance of our proposed technique and in the analysis, we consider an independent but not identically distributed Generalized-$K$ composite fading model. More specifically exact and closed-form expressions for the outage probability, the moment generating function, the moments of signal-to-noise ratio, and the average symbol error probability are derived. In addition, asymptotic outage probability and symbol error probability expressions are also presented in order to investigate the diversity order and the array gain. Finally, our theoretical performance results are validated by Monte Carlo simulations.

## I. INTRODUCTION

SIGNIFICANT improvement can be obtained in the performance of wireless communication systems when multiple antennas are used at the transmitter side [1], [2]. However, the employment of multiple antennas at transmitter increases the cost, complexity, and power consumption due to the increase in the number of required radio frequency (RF) chains [3], [4]. Transmit antenna selection (TAS) reduces the number of RF chains and also the cost, complexity, and power consumption since it relies on a single transmitter structure [3], [4]. In single TAS, a transmit antenna providing the highest instantaneous signal-to-noise ratio (SNR) at the receiver is selected and the selected antenna index is sent by the receiver to the transmitter over a feedback channel.

Because of the advantages, TAS has attracted great interest and has been for example studied in [5]–[10] for different receiver combining techniques. For instance, in [5]–[7], the performance of TAS has been investigated for Rayleigh fading channels. In [8], [9] and [10], the authors have studied the performance of TAS in Nakagami-$m$ fading, considering the independent identically distributed (*i.i.d.*) and independent but not identically distributed (*i.n.d.*) cases, respectively.

It should be noted that, in [5]–[10], the performance of the TAS scheme has been studied for a channel model including small-scale fading only. Moreover, in [5]–[10], the studied TAS scheme requires continuous estimation of the channel state information (CSI) coefficients of the branches at the receiver. These CSI coefficients and thus the index of the selected antenna change rapidly and as such a high rate feedback channel is required. This is typically an undesired situation because feedback channels have often low transmission rates and their numbers are limited.

In this work, we focus on a single transmit antenna selection method which is based on shadowing side information (SSI) [11]. Since SSI changes slowly as compared to CSI, by using TAS based on SSI the frequency of the usage of the feedback channel and the channel estimation complexity at the receiver can be reduced. In CSI based TAS, the receiver has to estimate the CSI of all transmit antennas in order to select the transmit antenna. On the other hand in the proposed technique, TAS is done based on SSI, and receiver has to estimate the CSI of only selected transmit antenna in order to use in maximum likelihood detection. This leads to a simpler receiver design. This technique can be used for the system involving multiple antenna transmitters which have a single RF chain. In addition, this technique can also be used in distributed transmit antenna systems in which a mobile station selects the best base station/access point [10]. Besides, the next generation wireless communications systems require higher data rate, which can be provided using millimeter wave communications (60 GHz and beyond). In higher frequency bands, shadowing effect comes into prominence [12], [13]. In addition, in these high frequency bands, CSI varies very fast and TAS based on instantaneous CSI becomes useless because the selected antenna index varies very fast. In this context, we analyze the performance of TAS based on SSI scheme in i.n.d. Generalized-$K$ fading channels. Generalized-$K$ composite fading model [14], [15] is a generalization of the $K$ distribution [16], [17].

The remainder of the paper is organized as follows. The system and channel models are described in Section II. In Section III, the outage probability, moments and moment generating function (MGF) and some symbol error probability (SEP) expressions are derived. In addition, asymptotical outage probability and SEP expressions are found to obtain the diversity order and the array gain of the TAS based on SSI. Next, numerical and simulation results are given in Section IV. Finally, conclusions are drawn in Section V.

⋆This work was supported in part by King Abdullah University of Science and Technology (KAUST) and conducted while Ahmet Yılmaz was visiting KAUST.

Ahmet Yılmaz and Oğuz Kucur are with the Department of Electronics Engineering, Gebze Institute of Technology, 41400, Gebze-Kocaeli, Turkey. e-mails: {ahmetyilmaz,okucur}@gyte.edu.tr.

Ferkan Yılmaz, and M.-S. Alouini are with the KAUST, Thuwal, Makkah Province, Saudi Arabia e-mails: {ferkan.yilmaz, slim.alouini}@kaust.edu.sa.

## II. System and Channel Models

We consider a wireless communication system in which the transmitter is equipped with $L$ antennas and the receiver is equipped with a single antenna. The links between the transmit antennas and the receiver are affected by an *i.n.d.* generalized-K composite fading. The receiver selects a single transmit antenna which provides the highest shadowing coefficient between the transmitter and receiver. The index of the selected transmit antenna is then sent by the receiver to the transmitter. In our analysis, we assume an error/delay free feedback channel.

Assuming that the $r^{th}$ transmit antenna is selected the conditional received SNR can be expressed as

$$\gamma_{end,r} = (E_s/N_0) h_r^2 \max_{1\leq\ell\leq L}\{g_\ell^2\} = \alpha_{max}\beta_r, \quad (1)$$

where $E_s$ is average energy per symbol, $N_0$ is one-sided power spectral density of the additive white Gaussian noise, $h_r$ is the fading coefficient between the selected antenna and receiver and $g_\ell$, $\ell = 1, 2, \ldots, L$ are the shadowing coefficients between transmit antennas and receiver. In (1) $\beta_r = h_r^2$, $\alpha_{max} = \max_{1\leq\ell\leq L}\{\alpha_\ell\}$, and $\alpha_\ell = (g_\ell^2 E_s/N_0)$, $\ell = 1, 2, \ldots, L$. Since we assume $g_\ell$s and $h_\ell$s are *i.n.d.* Nakagami-$m$ random variables, $\alpha_\ell$s and $\beta_\ell$s are Gamma random variables with the parameters $m_{\alpha,\ell}$ and $m_{\beta,\ell}$, $\mathbb{E}[\alpha_\ell] = \mathbb{E}[g_\ell^2](E_s/N_0) = (\Omega_\ell E_s/N_0)$ [1]. Without any loss of generality we assume that $\mathbb{E}[\beta_\ell] = 1$.

## III. Performance Analysis

### A. Outage Probability

Outage probability ($P_{out}$) is defined as the probability that the received SNR ($\gamma_{end}$), falls below a certain threshold value ($\gamma_{th}$) [18]. We can write $P_{out} = F_{\gamma_{end}}(\gamma_{th})$, where $F_{\gamma_{end}}(\cdot)$ is cumulative distribution function (CDF) of received SNR.

**Theorem 1.** *When the $r^{th}$ antenna is selected, the CDF of received SNR can be expressed as*

$$F_{\gamma_{end,r}}(x) = 1 + \frac{2}{\Gamma(m_{\beta,r})}\sum\sum \kappa_{n,k}\sqrt{\frac{(xm_{\beta,r})^{m_{\beta,r}+A_{n,k}}}{B_n^{A_{n,k}-m_{\beta,r}}}}$$
$$\times K_{m_{\beta,r}-A_{n,k}}\left(2\sqrt{xB_n m_{\beta,r}}\right), \quad (2)$$

*where*

$$\kappa_{n,k} = \prod_{\ell=1}^{L} \frac{(-1)^{n_\ell} m_{\alpha,\ell}^{n_\ell(k_\ell+1)-1}}{(\bar{\gamma}_\ell^{k_\ell} k_\ell!)^{n_\ell}},$$

$$B_n = \sum_{\ell=1}^{L} n_\ell \frac{m_{\alpha,\ell}}{\bar{\gamma}_\ell}, \qquad A_{n,k} = \sum_{\ell=1}^{L} k_\ell n_\ell. \quad (3)$$

*Proof:* Proof of Theorem 1 is given in Appendix A. ∎

In (2), $\bar{\gamma}_\ell = \mathbb{E}[\alpha_\ell]$, $\Gamma(\cdot)$ is the Gamma function [19, Eq. (8.310.1)] and $K_v(\cdot)$ is the $v$th order modified Bessel function of second kind [19, Eq.(8.407.1)]. As a short hand notation, the following notation: $\sum\sum \equiv \sum_{n\in\theta_L}\sum_{k_1=0}^{m_{\alpha,1}-1}\cdots\sum_{k_L=0}^{m_{\alpha,L}-1}$, is used in (2) and will be used from now on.

---

[1] In this paper, $\mathbb{E}[\cdot]$ denotes the expectation operator.

The outage probability of the received SNR for any selected antenna can be obtained by averaging (2) over antenna selection probabilities yielding

$$F_{\gamma_{end}}(x) = \sum_{r=1}^{L} P_r F_{\gamma_{end,r}}(x), \quad (4)$$

where $P_r$ is the probability of $r^{th}$ antenna selection. If the $r^{th}$ antenna is selected, the shadowing coefficient of the $r^{th}$ antenna is the largest among all other coefficients. Let $X = X_1, X_2, \ldots, X_{L-1}$ be the set of shadowing coefficients except the selected one ($\alpha_{max}$) which is bigger than the maximum of $X$. As such, we can write the following equation for $P_r$

$$P_r = P(\max\{X_1, X_2, \ldots, X_{L-1}\} \leq X_{max}|\alpha_r)$$
$$= \int_0^\infty F_{X,max}(x) f_{\alpha,r}(x) dx, \quad (5)$$

where $F_{X,max}(\cdot)$ is the CDF of the maximum of $L-1$ Gamma random variables. Substituting (A.7) with $L-1$ and $f_{\alpha,r}(x) = (m_{\alpha,r}/\Omega_{\alpha,r})^{m_{\alpha,r}} x^{m_{\alpha,r}-1}\exp(-xm_{\alpha,r}/\Omega_{\alpha,r})/\Gamma(m_{\alpha,r})$ into (5), and using [19, Eq.(3.326)] we obtain after some manipulations

$$P_r = \frac{1}{\Gamma(m_{\alpha,r})}\left(\frac{m_{\alpha,r}}{\Omega_{\alpha,r}}\right)^{m_{\alpha,r}} \sum_{n\in\theta_{L-1}} \sum_{k_1=0}^{m_{\alpha,1}-1}\cdots\sum_{k_{L-1}=0}^{m_{\alpha,L-1}-1}$$
$$\times \frac{\kappa_{n,k}\Gamma(A_{n,k}+m_{\alpha,r})}{(B_n+m_{\alpha,r}/\Omega_{\alpha,r})^{A_{n,k}+m_{\alpha,r}}}. \quad (6)$$

### B. Moments of the Received SNR

By using the first two moments of the received SNR, important performance measures such as average received SNR and the amount of fading (AF) can be obtained. AF is defined as $AF_{\gamma_{end}} = \left(\mathbb{E}_{\gamma_{end}}[x^2] - (\mathbb{E}_{\gamma_{end}}[x])^2\right)/(\mathbb{E}_{\gamma_{end}}[x])^2$ [18]. The $p$th order moment of the received SNR can be derived by following integral $\mathbb{E}_{\gamma_{end}}[x^p] = p\int_0^\infty x^{p-1}[1-F_{\gamma_{end}}(x)]dx$. If (4) is substituted and [19, Eq.(7.811.4)] is used, we obtain the following closed-form expression for the moments

$$\mathbb{E}_{\gamma_{end}}[x^p] = -p\sum_{r=1}^{L}\sum\sum \frac{P_r \kappa_{n,k}\Gamma\left(p+m_{\beta,r}+\frac{A_{n,k}}{2}\right)}{\Gamma(m_{\beta,r})}$$
$$\times \frac{\Gamma\left(p+A_{n,k}+\frac{m_{\beta,r}}{2}\right)}{\sqrt{m_{\beta,r}^{A_{n,k}+m_{\beta,r}+2p} B_n^{2A_{n,k}+m_{\beta,r}+2p}}}. \quad (7)$$

### C. Moment Generating Function

The MGF is a well known tool to obtain performance of wireless communication systems [18]. The MGF can be derived by using the CDF as $\mathcal{M}_{\gamma_{end}}(s) = s\int_0^\infty e^{-sx} F_{\gamma_{end}}(x) dx$. By substituting the CDF given in (4) and using [19, Eq.(6.643.3)], the following closed-form expression is obtained for the MGF

$$\mathcal{M}_{\gamma_{end}}(s) = 1 + \sum_{r=1}^{L}\sum\sum \frac{2P_r \kappa_{n,k} m_{\beta,r}\Gamma(A_{n,k}+1)}{\sqrt{e^{-\mu}/\mu^{m_{\beta,r}+A_{n,k}-1}}B_n^{A_{n,k}}}$$
$$\times \frac{W_{-\frac{m_{\beta,r}+A_{n,k}+1}{2},\frac{m_{\beta,r}-A_{n,k}}{2}}(\mu)}{\Gamma(m_{\beta,r})}, \quad (8)$$

where $\mu = B_n m_{\beta,r}/s$ and $W_{a,b}(\cdot)$ is the Whittaker-W function which is defined in [19, Eq.(9.222)].

### D. Symbol Error Probability

By using the MGF based approach, the performance of a wide variety of modulations can be expressed in a single integral representation [18]. However, in this section we give closed-form SEP expressions for modulations which have a conditional SEP in the form

$$P_s = \mathbb{E}_{\gamma_{end}}\left[a\,Q\left(\sqrt{2b\gamma_{end}}\right)\right], \quad (9)$$

where $Q(\cdot)$ is the Gaussian $Q$-function defined by [18, Eq. (4.1)], and the coefficients $a$ and $b$ are modulation dependent constants. (9) provides the exact SEP results for binary phase shift keying (BPSK) ($a = 1$, $b = 1$), orthogonal binary frequency shift keying ($a = 1$, $b = 0.5$), and $M$-ary pulse amplitude modulation [$a = 2(M-1)/M$, $b = 3/(M^2-1)$]. Moreover, (9) provides approximate SEP results for $M$-PSK [$a = 2$, $b = \sin^2(\pi/M)$] and rectangular $M$-ary quadrature amplitude modulation [$a = 4 - 4/\sqrt{M}$, $b = 1.5/(M-1)$] [22]. When (9) is rewritten in terms of CDF, the following SEP is obtained $P_s = \frac{a\sqrt{b}}{2\sqrt{\pi}}\int_0^\infty \frac{e^{-bx}}{\sqrt{x}} F_{\gamma_{end}}(x)dx$, by substituting the CDF given in (4) and then using [19, Eq.(6.643.3)], a closed-form expression for the SEP is obtained as

$$P_s = \frac{a}{2} + \sum_{r=1}^{L}\sum\sum \frac{P_r a \kappa_{n,k}\Gamma\left(A_{n,k}+\frac{1}{2}\right)\Gamma\left(m_{\beta,r}+\frac{1}{2}\right)}{2B_n^{A_{n,k}}\Gamma(m_{\beta,r})\sqrt{\pi}e^{-\xi/\xi^{m_{\beta,r}+A_{n,k}-1}}}$$
$$\times W_{-\frac{m_{\beta,r}+A_{n,k}}{2},\frac{m_{\beta,r}-A_{n,k}}{2}}(\xi), \quad (10)$$

where $\xi = B_n m_{\beta,r}/b$.

### E. Asymptotical Performance Analysis

The analytical performance expressions derived in previous sections do not reveal any information about the diversity order of the network. Therefore, in order to provide further insight on our performance analysis and obtain the diversity order of the TAS scheme based on SSI, we use the high SNR approximation technique presented in [23] to derive asymptotical outage probability and SEP expressions.

**Theorem 2.** *The outage probability of the TAS scheme based on SSI can asymptotically be expressed as:*

$$P_{out} \approx (\zeta/d)(\gamma_{th}/\bar{\gamma})^d + \mathcal{O}\left(\bar{\gamma}^{-d}\right) \quad (11)$$

*where $\bar{\gamma} = \kappa_1\bar{\gamma}_1 = \kappa_2\bar{\gamma}_2 = \cdots = \kappa_L\bar{\gamma}_L$, $\kappa_\ell$, $\ell = 1,\ldots,L$, are positive constants and $d = \min(d_\alpha, d_\beta)$, $d_\alpha = \sum_{\ell=1}^{L} m_{\alpha,\ell}$, $d_\beta = \min_{1 \leq r \leq L}(m_{\beta,r})$. In (11), $\zeta$ is given as follows:*

$$\zeta = \begin{cases} \sum_{j=1}^{L} \frac{P_\beta Z_L \Gamma(\Delta_d)(\kappa_j)^{\Delta_d}}{\Gamma(d_\beta) d_\beta^{-d_\beta}(m_{\alpha,j})^{\Delta_d-1}}, & d_\alpha > d_\beta, \\ \sum_{j=1}^{L} \frac{m_{\alpha,j} Z_L \ln\left(\frac{\bar{\gamma}_j \psi d_\beta^{-1}}{xm_{\alpha,j}}\right)}{d_\beta^{-d_\alpha}\Gamma(d_\beta)/P_\beta}, & d_\alpha = d_\beta, \quad (12)\\ \frac{P_\beta L Z_L \Gamma(-\Delta_d)(d_\beta)^{d_\alpha}}{\Gamma(d_\beta)}, & d_\alpha < d_\beta, \end{cases}$$

where $\Delta_d = d_\alpha - d_\beta$, $\psi$ denotes Euler's constant [24, Eq. (1.7.7)], $Z_L = \prod_{\ell=1}^{L} \frac{(m_{\alpha,\ell}/\kappa_\ell)^{m_{\alpha,\ell}}}{\Gamma(m_{\alpha,\ell}+1)}$ and $P_\beta$ is the selection probability of the antenna whose instantaneous SNR distribution has the smallest $m_{\beta,r}$.

Asymptotic SEP expression of TAS based on SSI for the modulations, which have a conditional SEP given in (9), can also be obtained as

$$P_s \approx 2^{d-1} a\zeta\Gamma(d+1/2)\left(d\sqrt{\pi}\right)^{-1}(2b\bar{\gamma})^{-d} + \mathcal{O}\left(\bar{\gamma}^{-d}\right) \quad (13)$$

*Proof:* The proof of Theorem 2 is given in Appendix B. ∎

Since SEP is obtained asymptotically as $P_s \approx (G_a\bar{\gamma})^{-G_d}$, where $G_a$ and $G_d$ denote array gain and diversity gain (order), respectively, we can deduce the array gain expression as $G_a = 2b\left(2^{d-1}a\zeta\Gamma(d+1/2)/(d\sqrt{\pi})\right)^{-1/d}$, and the diversity order expression as

$$G_d = \min\{d_\alpha, d_\beta\}. \quad (14)$$

## IV. NUMERICAL RESULTS

In this section, we give some numerical results investigating the performance of TAS based on SSI. Numerical results, which are obtained based on our theoretical results are supported by Monte Carlo simulations, in which $N$ repeated random experiments based on the system model are executed and then the number of occurrences of the interested event is counted and divided by $N$, which gives the estimation for the probability of the interested event. In this work, $N$ is taken to be 20 million. In the numerical results, $m_\alpha = 1$ and $m_\alpha = 3$ are used for heavy shadowing and light shadowing cases, respectively [25].

Figure 1 depicts the outage probability performance of the TAS scheme based on SSI for different number of antennas, *i.i.d.* channels and $m_{\alpha,\ell} = m_\alpha$ & $m_{\beta,\ell} = m_\beta$, $\ell = 1, 2, \ldots, L$. In Figure 1, the curve with $L = 1$ shows the outage probability performance of single transmit antenna system. As seen in Figure 1, as the number of antennas at the transmitter increases outage probability performance of the system also increases. Moreover, for $m_\alpha = m_\beta = 1$, the outage probability performances of the TAS scheme based on SSI for $L = 2$, $L = 3$ and $L = 4$ are 9 dB, 11 dB and 12 dB better than that of $L = 1$ case at $P_{out} = 10^{-4}$, respectively. Since in TAS based on SSI, the transmit antenna is selected with respect to shadowing coefficients, the fading coefficient of the selected antenna ends up being assigned randomly. Therefore, the presented result in Figure 1, illustrates performance of the systems limited (determined) by statistical properties of fading coefficients. Since $\sum_{\ell=1}^{L} m_{\alpha,\ell} \geq \min_{1 \leq r \leq L}(m_{\beta,r})$, the negative slopes of the curves (i.e., diversity orders) for TAS based on SSI curves are the same as the curve for $L = 1$. This means that only an array gain can be obtained in this case. The performance improvement of the TAS based on SSI is much more in heavy shadowing case. Since the antenna selection is done based on shadowing coefficient, TAS based on SSI provides more gain when the channel has severe shadowing. Also as seen in Figure 1, the analytical and simulation results are in perfect agreement.

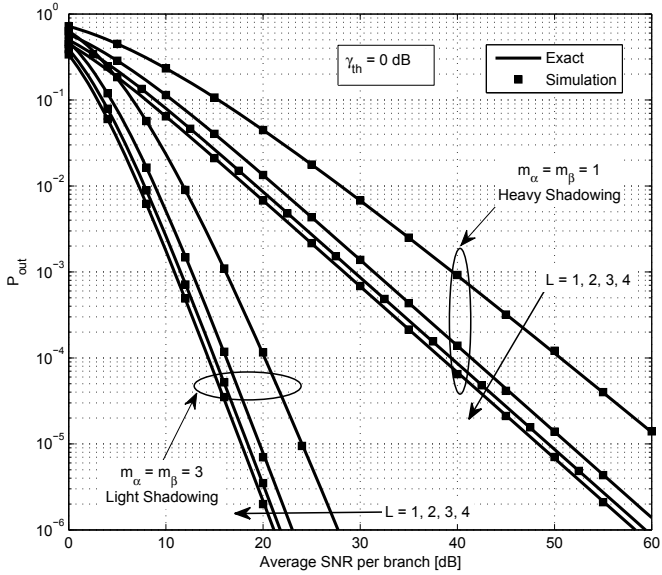

Fig. 1. Outage probability results for TAS based on shadowing for different number of antennas and *i.i.d.* channels.

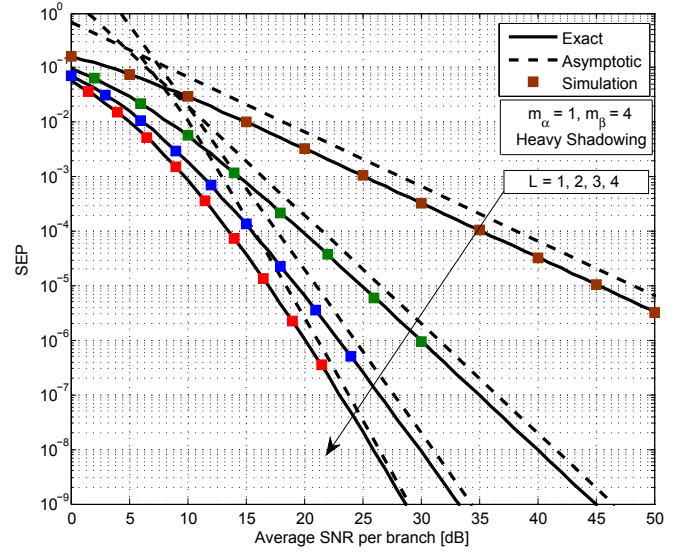

Fig. 2. SEP results for TAS based on shadowing for different number of antennas, *i.i.d.* channels and BPSK modulation.

In Figure 2, SEP performance results of TAS based on SSI are presented for BPSK modulation, different number of antennas and *i.i.d.* channels $m_{\alpha,\ell} = m_\alpha = 1$ & $m_{\beta,\ell} = m_\beta = 4$, $\ell = 1, 2, \ldots, L$. As shown in Figure 2, TAS scheme based on SSI not only provides an array gain as in the cases presented in Figure 1 but also provides a certain diversity gain. For the presented results since shadowing coefficients (with $m_{\alpha,\ell} = m_\alpha = 1$, $\ell = 1, 2, \ldots, L$) dominate fading coefficients (with $m_{\beta,\ell} = m_\beta = 4$, $\ell = 1, 2, \ldots, L$), then performance of the system is limited by the shadowing coefficients and by selecting transmit antenna based on shadowing coefficients (i.e., SSI) diversity gain can be obtained. For the presented results, diversity order of the system is equal to 2, 3 and 4 for $L = 2$, $L = 3$ and $L = 4$, respectively. Moreover, negative slopes of the exact curves and asymptotical curves are very compatible especially in the high SNR regime.

Figure 3 presents SEP performance of BPSK modulation over *i.n.d.* channels, for $L = 3$, $m_\alpha = \{m_{\alpha,1}, m_{\alpha,2}, m_{\alpha,3}\} = \{1, 1, 1\}$, $m_\beta = \{m_{\beta,1}, m_{\beta,2}, m_{\beta,3}\} = \{2, 3, 1\}$ and $L = 2$, $m_\alpha = \{m_{\alpha,1}, m_{\alpha,2}\} = \{3, 2\}$, $m_\beta = \{m_{\beta,1}, m_{\beta,2}\} = \{2, 3\}$. In Figure 3, we compare TAS based on SSI to random TAS without shadowing where single transmit antenna is selected randomly and shadowing effect is ignored. As seen in Figure 3, SEP performance of TAS based on SSI is slightly better or worse than that of random TAS without shadowing. This means that by using TAS based on SSI, the shadowing effects on the transmitted signal are eliminated or enlightened. Moreover, it is clearly seen that the diversity orders of the curves are as given in (14).

In Figure 4, the SEP performance results are presented in the presence of correlation between shadowing coefficients In the simulations, power correlation model is used [18]. According to results as the correlation increases between shadowing coefficients the gain coming along with TAS based on SSI reduces.

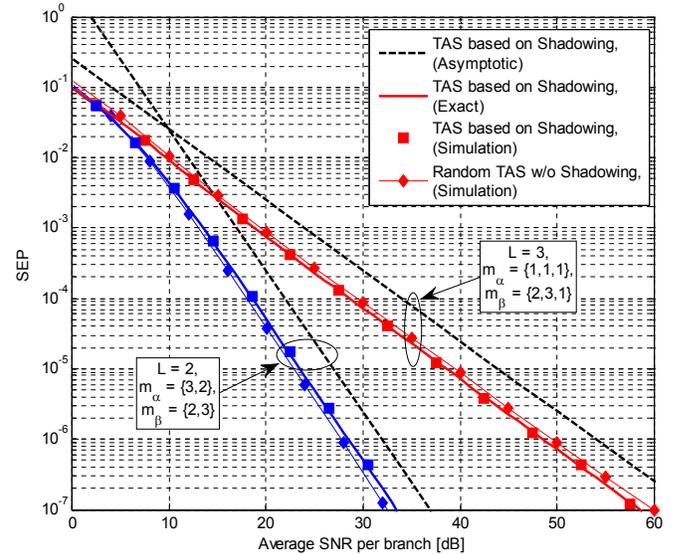

Fig. 3. SEP results for TAS based on shadowing for different number of antennas, *i.n.d.* channels and BPSK modulation.

Figure 5 illustrates the impact of the feedback error on the SEP performance of the TAS based on SSI for different channel conditions and probability of feedback error ($p_e$). In order to simulate the feedback error case we use the same feedback channel model described in [26], [27]. In this model, index of the selected antenna is sent by using binary digits and antenna index requires $k = \lceil \log_2(L) \rceil$ bits, where $L$ is the number of transmit antennas and $\lceil x \rceil$ denotes nearest integer which is greater than or equal to $x$. The feedback channel is modeled as binary symmetric channel; thus probability of feedback error ($p_e$) is the same for each feedback bit. Index of the selected antenna is received correctly and incorrectly with the probability $P_{cf} = (1 - p_e)^k$ and $P_{ef} = 1 - P_{cf}$, respectively. According to results in Figure 5, in the presence of feedback error performance of the TAS based on SSI degrades. However, as the probability of feedback error reduces performance of the system improves.

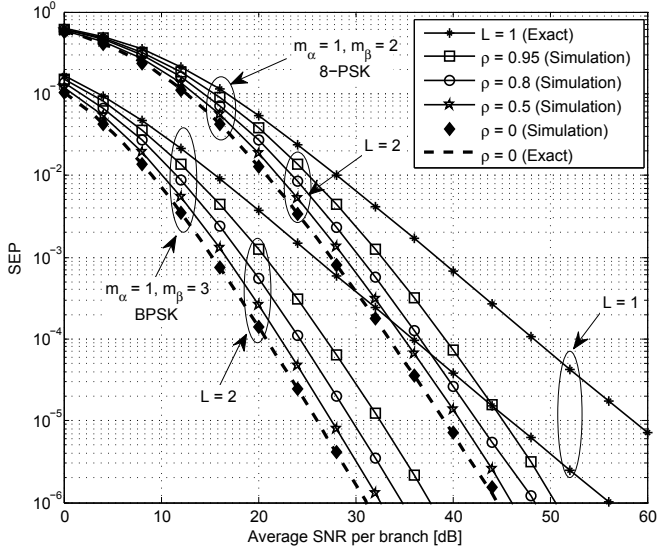

Fig. 4. Impact of the correlation between shadowing coefficients on the performance of the TAS based on SSI for different fading parameters and different correlation coefficients.

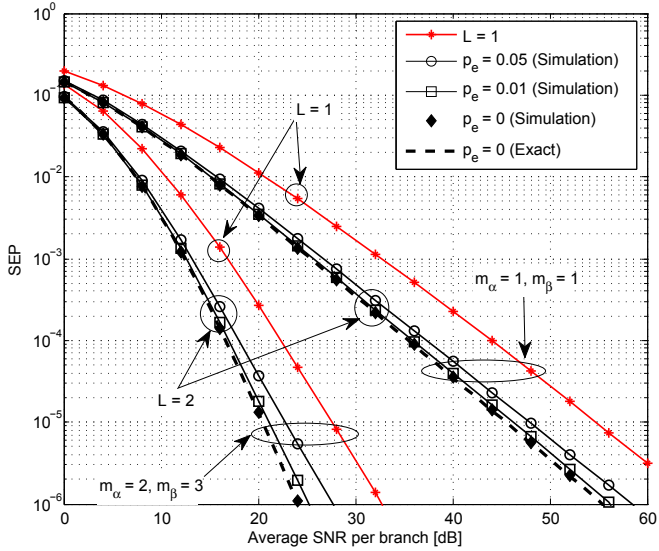

Fig. 5. Impact of the feedback error on the performance of the TAS based on SSI for different fading and shadowing parameters and different probability of feedback errors and BPSK modulation.

## V. CONCLUSIONS

In this work, a new transmit antenna selection scheme has been proposed. In this scheme, a single transmit antenna is selected based on the shadowing side information. Exact closed-form outage probability, moments, MGF and SEP expressions have been derived for the generalized-$K$ fading composite channel model. In addition, the diversity order and array gain of the system have been obtained by deriving asymptotical outage probability and SEP expressions. According to the obtained results, as the number of antennas increases, the performance of the TAS based on SSI improves and the shadowing effect on transmitted signal can be eliminated. In addition, the diversity order of the system is equal to minimum of sum of shadowing parameters ($m_{\alpha,\ell}$, $\ell = 1, 2, \ldots, L$) and minimum of fading parameters ($m_{\beta,\ell}$, $\ell = 1, 2, \ldots, L$).

## APPENDIX A

*Proof of Theorem 1:* The CDF of the conditional received SNR given in (1) can be obtained as follows

$$F_{\gamma_{end,r}}(x) = \int_0^\infty F_{\beta,r}(x|u) f_{\alpha,max}(u) du, \quad (A.1)$$

where $F_{\beta,r}(x|u)$ is the conditional CDF of SNR for the $r$th branch which is given by $F_{\beta,r}(x|u) = \gamma(m_{\beta,r}, x m_{\beta,r}/u)/\Gamma(m_{\beta,r})$, where $\gamma(\cdot, \cdot)$ is the lower incomplete Gamma function [19, Eq. (8.350.1)]. Besides, by using [21, Eq.(07.34.03.0088.01)] and [19, Eq.(9.31.5)], it can also be expressed in terms of the Meijer-G function [19, Eq.(9.301)] as

$$F_{\beta,r}(x|u) = 1 - \frac{G_{2,1}^{0,2}\left[\frac{u}{xm_{\beta,r}} \bigg| \begin{array}{c} 1, 1-m_{\beta,r} \\ 0 \end{array}\right]}{\Gamma(m_{\beta,r})}. \quad (A.2)$$

In (A.1), $f_{\alpha,max}(\cdot)$ is PDF of $\alpha_{max}$. Since the $\alpha_\ell$s have a Gamma distribution, their CDF is expressed as $F_{\alpha,\ell}(x) = \gamma(m_{\alpha,\ell}, x m_{\alpha,\ell}/\bar{\gamma}_\ell)/\Gamma(m_{\alpha,\ell})$. By using the property of the lower incomplete Gamma function given in [19, Eq. (8.352.6)], $F_{\alpha,\ell}(x)$ can be expressed as

$$F_{\alpha,\ell}(x) = 1 - e^{-x\frac{m_{\alpha,\ell}}{\bar{\gamma}_\ell}} \sum_{k=0}^{m_{\alpha,\ell}-1} \frac{1}{k!} \left(x \frac{m_{\alpha,\ell}}{\bar{\gamma}_\ell}\right)^k, \quad (A.3)$$

when the $m_{\alpha,\ell}$s are restricted to be integer values. Since $\alpha_{max}$ is the highest one among $\alpha_\ell$, $\ell = 1, 2, \ldots, L$, we can write the CDF of $\alpha_{max}$ in terms of CDFs of $\alpha_\ell$s as, $F_{\alpha,max}(x) = \prod_{l=1}^{L} F_{\alpha,\ell}(x)$ [20]. If (A.3) is substituted we obtain

$$F_{\alpha,max}(x) = \prod_{\ell=1}^{L} \left\{ 1 - e^{-\frac{xm_{\alpha,\ell}}{\bar{\gamma}_\ell}} \sum_{k_\ell=0}^{m_{\alpha,\ell}-1} \frac{(xm_{\alpha,\ell})^{k_\ell}}{\bar{\gamma}_\ell^{k_\ell} k_\ell!} \right\}. \quad (A.4)$$

**Lemma.** *Let $g_\ell(x)$ be an arbitrary function, $\theta_L$ is the set of binary numbers whose lengths are $L$ and $n_\ell$ is the $\ell^{th}$ element of binary number $\mathbf{n}$. Note that one can show that the following equation is valid*

$$\prod_{\ell=1}^{L} \{1 - g_\ell(x)\} = \sum_{\mathbf{n} \in \theta_L} \prod_{\ell=1}^{L} (-1)^{n_\ell} [g_\ell(x)]^{n_\ell}. \quad (A.5)$$

*Proof of Lemma:* In order to prove the Lemma, we give some illustrative examples. Let $x_l$ be an arbitrary variable or arbitrary function, we can write following equation:

$$\prod_{\ell=1}^{2} (1 - x_\ell) = (1 - x_1)(1 - x_2)$$
$$= x_1^0 x_2^0 - x_1^0 x_2^1 - x_1^1 x_2^0 + x_1^1 x_2^1.$$

In general, one can show that powers of $x_\ell$s are the binary numbers and they are elements of the set including binary numbers whose lengths are upper limit of production. Moreover, signs of the terms are minus or plus when sum of the powers is an odd number or even number, respectively. In this way we can write a general expression as given in (A.5) for the product given in (A.4). ∎

By using our novel representation given in Lemma above we can express $F_{\alpha,max}(x)$ as

$$F_{\alpha,max}(x) = \sum_{\boldsymbol{n}\in\theta_L} \prod_{\ell=1}^{L} \left[ -e^{-\frac{xm_{\alpha,\ell}}{\bar{\gamma}_\ell}} \sum_{k_\ell=0}^{m_{\alpha,\ell}-1} \frac{(xm_{\alpha,\ell})^{k_\ell}}{\bar{\gamma}_\ell^{k_\ell} k_\ell!} \right]^{n_\ell} \quad \text{(A.6)}$$

Since $n_\ell \in \{0,1\}$, (A.6) can be expressed as

$$F_{\alpha,max}(x) = \sum\!\!\!\sum \kappa_{n,k} e^{-xB_n} x^{A_{n,k}}, \quad \text{(A.7)}$$

where $\kappa_{n,k}$, $B_n$ and $A_{n,k}$ are as given in (3). The PDF of maximum of i.n.d. Gamma random variables can be obtained by taking the derivative of (A.7), and can be written as $f_{\alpha,max}(x) = \sum\!\!\!\sum \kappa_{n,k} e^{-xB_n} x^{A_{n,k}-1}(A_{n,k} - xB_n)$ and can also be expressed by using [21, Eq.(07.34.03.0228.01) and Eq.(07.34.20.0001.01)] in terms of the Meijer-G function as

$$f_{\alpha,max}(x) = \sum\!\!\!\sum \frac{\kappa_{n,k}}{B_n^{A_{n,k}-1}} G_{1,1}^{1,2}\!\left[ xB_n \,\bigg|\, \begin{matrix} -1 \\ A_{n,k}-1,\ 0 \end{matrix} \right]. \quad \text{(A.8)}$$

Using properties of the Meijer-G function given by [19, Eq. (7.811.1) and Eq. (9.31.1)], $F_{\gamma_{end,r}}(x)$ in (A.1) can be derived as

$$F_{\gamma_{end,r}}(x) = 1 - \sum\!\!\!\sum \frac{\kappa_{n,k} G_{0,2}^{2,0}\!\left[ xB_n m_{\beta,r} \,\bigg|\, \begin{matrix} - \\ m_{\beta,r},\ A_{n,k} \end{matrix} \right]}{\Gamma(m_{\beta,r}) B_n^{A_{n,k}}}. \quad \text{(A.9)}$$

Using [19, Eq.(9.31.1) and Eq.(9.34.3)], the Meijer-G function in (A.9) reduces to a modified Bessel function of second kind, and then $F_{\gamma_{end,r}}(x)$ can be written as given in (2). ∎

## APPENDIX B

*Proof of Theorem 2:* Using the proposed method in [23] to obtain asymptotic outage probability and SEP expressions as given in (11) and (13), respectively, we have first to express the PDF of the received SNR as $f_{\gamma_{end}}(x) \approx \zeta(x/\bar{\gamma})^d + \mathcal{O}(\bar{\gamma}^{-d})$. The PDF of received SNR can be obtained as

$$f_{\gamma_{end}}(x) = \sum_{r=1}^{L} \int_0^\infty f_{\beta,r}(x|u) f_{\alpha,max}(u) du, \quad \text{(B.1)}$$

where

$$f_{\beta,r}(x|u) = \left(\frac{m_{\beta,r}}{u}\right)^{m_{\beta,r}} \frac{x^{m_{\beta,r}-1}}{\Gamma(m_{\beta,r})} e^{-x\frac{m_{\beta,r}}{u}}, \quad \text{(B.2)}$$

and

$$f_{\alpha,max}(x) = \sum_{j=1}^{L} f_{\alpha,j}(x) \prod_{\substack{\ell=1 \\ \ell\neq j}}^{L} F_{\alpha,\ell}(x). \quad \text{(B.3)}$$

We can asymptotically express the $F_{\alpha,\ell}(x)$ by using the property of lower incomplete gamma function given by [24, Eq. (45:9:1)] as $F_{\alpha,\ell}(x) \approx (xm_{\alpha,\ell}/\bar{\gamma}_\ell)^{m_{\alpha,\ell}}/\Gamma(m_{\alpha,\ell}+1)$.

If we substitute approximate $F_{\alpha,\ell}(x)$ expression into (B.3) together with $f_{\alpha,j}(x) = \left(\frac{m_{\alpha,j}}{\bar{\gamma}_j}\right)^{m_{\alpha,j}} \frac{x^{m_{\alpha,j}-1}}{\Gamma(m_{\alpha,j})} e^{-x\frac{m_{\alpha,j}}{\bar{\gamma}_j}}$, we obtain after some manipulation

$$f_{\alpha,max}(x) \approx Z_L x^{-1+d_\alpha} \sum_{j=1}^{L} m_{\alpha,j} e^{-xm_{\alpha,j}/\bar{\gamma}_j}. \quad \text{(B.4)}$$

If (B.2) and (B.4) are substituted into (B.1) then by the help of [20, Eq. (3.471.9)] and after some manipulatons we obtain

$$f_{\gamma_{end}}(x) \approx \sum_{r=1}^{L}\sum_{j=1}^{L} \frac{\sqrt{(xm_{\beta,r})^{d_\alpha+m_{\beta,r}} (m_{\alpha,j}/\bar{\gamma}_j)^{m_{\beta,r}-d_\alpha}}}{x\Gamma(m_{\beta,r})/(2P_r Z_L)}$$

$$\times m_{\alpha,j} K_{d_\alpha - m_{\beta,r}}\!\left( \sqrt{\frac{xm_{\alpha,j} m_{\beta,r}}{\bar{\gamma}_j/4}} \right). \quad \text{(B.5)}$$

By using the asymptotic properties of the modified Bessel function of second kind given in [24, Sect. (50:9)], we can express the $f_{\gamma_{end}}(x)$ after several manipulations $f_{\gamma_{end}}(x) \approx \zeta(x/\bar{\gamma})^d + \mathcal{O}(\bar{\gamma}^{-d})$ where $d = \min(d_\alpha, d_\beta)$ and $\zeta$ is as given in (12). Consequently, we can write the asymptotic outage probability and SEP expressions given by (11) and (13), respectively. ∎